\documentstyle[aaspp4]{article}

\def\kms{km s$^{-1}$}
\def\etal{{\it et al.\ }}
\def\eg{{\it e.g.\ }}

\begin{document}

\title{Regular Orbits and Periodic Loops in Multiply--Barred Galactic 
Potentials}

\author{Witold Maciejewski and Linda S. Sparke}

\affil{Department of Astronomy, University of  Wisconsin, 475 North 
Charter Street, Madison, WI 53706-1582; witold@uwast.astro.wisc.edu, 
sparke@madraf.astro.wisc.edu}

\begin{abstract}
  We show that non-chaotic multiply-periodic particle orbits can exist
  in a galaxy-like potential where a small fast-tumbling nuclear bar
  is nested inside a main bar which has a slower pattern speed.  We
  introduce the concept of a {\it loop:} a one-dimensional curve such
  that particles distributed along it at some initial instant return
  to the same curve (as viewed in the rotating frame of one of the
  bars) after the bars return to the same relative position.  Just as
  particles following regular orbits in a simply-barred potential are
  trapped around closed periodic orbits, so regular orbits in a
  doubly-barred potential oscillate about stable loops. We find both
  loops which remain aligned with the inner bar and loops aligned with
  the outer bar: particles trapped around these loops could form the
  building blocks for a long-lived, self-consistent, doubly-barred
  galaxy.  In a realistic doubly-barred galaxy potential, we
  demonstrate the existence of stable loops which support the
  structure of the inner bar.  We use the epicyclic approximation to
  preview the pattern of gas flow in a doubly-barred potential.
\end{abstract}


\section{INTRODUCTION}
The idea of `bars-within-bars', with a small nuclear bar lying inside
a slower-rotating kiloparsec-scale bar, was introduced to explain how
gas could flow into the center of a galaxy with an active nucleus, to
feed `the monster' lurking there (\eg Shlosman, Frank \& Begelman
1989).  Torques due to a single bar can efficiently remove angular
momentum from gas and cause it to move inwards; but if the bar has an
inner Lindblad resonance, the gas accumulates in this region, which in
a kiloparsec-scale bar is generally at least 100 pc out, and flow into
the center is slow (\eg Piner \etal 1995).  Nuclear bars nested within
the main bar and having a whole range of position angles relative to
it have been observed in the optical and near IR (see Shaw \etal 1993;
1995; Wozniak \etal 1995; Friedli 1996; Erwin \etal 1997).
Self-consistent N-body simulations including a gas component (Friedli
\& Martinet 1993; Heller \& Shlosman 1994; Combes 1994) have produced
double bars which persist for at least several rotation periods and
significantly influence the gas distribution in the inner parts of the
galaxy.

Because there is no reference frame in which the potential of a
doubly-barred system is constant, the particle motions are often
assumed to be chaotic: we show here that this is a misconception;
these potentials admit families of regular, multiply-periodic orbits.
Our work runs somewhat parallel to that of Louis \& Gerhard (1988) and
of Sridhar (1989), who studied orbits in a pulsating spherical system.

\section{THE CONCEPT OF A LOOP}
In the symmetry plane of a single steadily-rotating bar, there are two
principal families of stable orbits which close in the rotating frame
of the bar, and lie within the corotation radius, at which the pattern
speed of the rotating bar is equal to the orbital period of the disk
stars: they are the $x_1$ and $x_2$ orbits, elongated respectively
parallel and perpendicular to the bar. In a self-consistent bar, most
of the stars will be on orbits trapped around the $x_1$ family (\eg
Contopoulos \& Papayannopoulos 1980, Sparke \& Sellwood 1987). On
these periodic orbits, only the action associated with a star's orbit
around the center is non-zero, and the motion is simply periodic.

Choosing spherical polar coordinates $R,\theta$ in the plane of the barred
disk, we can approximate the potential $\Phi (R,\theta)$ of a weakly barred 
galaxy as the sum of an axisymmetric part $\Phi_0(R)$ and a bisymmetric term 
\begin{equation}
\Phi_1 (R, \theta) = \Phi_B(R) \cos[2 (\theta-\Omega_B t) ] \/,
\end{equation}
where $\Omega_B$ is the bar's angular velocity. The motion of a test particle 
in this potential can be decomposed into the `guiding center' motion, on 
a circle of radius $R_0$ with the  angular velocity $\Omega(R_0)$ 
corresponding to the axisymmetric 
potential $\Phi_0$, together with the epicyclic oscillations resulting from 
the forcing term $\Phi_1$, and a free oscillation at the local epicyclic 
frequency $\kappa (R_0)$. On a periodic orbit, this last oscillation is absent, 
and in the rotating frame of the bar we can write the deviations $R_1, \theta_1$
from a circular orbit as 
\begin{equation}
R_1 = B_R \cos (\omega t + 2 \theta_S) \/; \hspace{5mm} 
\theta_1 = B_{\theta} \sin (\omega t + 2 \theta_S) \/,
\end{equation}
where the frequency $\omega = 2 [\Omega(R_0)-\Omega_B]$, and the constants
 $B_R, B_{\theta}$ are functions of the potential; 
$\theta_S$ gives the initial position of the particle
(see \eg   Section 3.3.3 of Binney \& Tremaine 1987). 
Since the particle returns to the same point after a time $2\pi/\omega$, 
this orbit is closed.

A second smaller bar would correspond to an additional potential of the
form (1), where  $\Phi_B$ and $\Omega_B$ are replaced by the quantities
$\Phi_S$ and $\Omega_S$ appropriate to the second bar.
Since the equations of motion are linear, the new forced oscillation is 
the sum of those induced by the two bar potentials separately;
in the frame rotating with the larger bar, 
we can write the displacement in radius as
\begin{equation}
R_1 = B_R \cos [\omega t + 2 \theta_S ] + 
S_R \cos [ (\omega-\omega_p) t + 2 \theta_S ] \/, 
\end{equation}
where $\omega_p = 2(\Omega_S-\Omega_B)$, with a similar equation for
 $\theta_1$. Here $\theta_S$ is the 
initial angular coordinate of the particle, where we measure the time $t$ 
from the moment at which the two bars are aligned. {\it 
We define a loop as the one-dimensional curve defined by varying $\theta_S$:
if at any instant one places particles along this curve,
then after a time $2\pi/\omega_p$, they will 
all have returned to the curve, but will 
have moved around it by an angle $2 \pi \omega / \omega_p$.}
In this linear approximation, a loop is a set of particles 
which have the same guiding radius, and lack any free epicyclic motion,
but respond only to the periodic forcing of the two bars;
a general orbit with that same guiding center will oscillate about the loop.

Abandoning the epicyclic approximation, the two-dimensional problem of
motion in the symmetry plane of a general doubly-barred system can be
reformulated as that of motion in a three-dimensional time-steady
potential (Lichtenberg \& Lieberman 1992, p.15). A regular orbit in
such a potential can be expressed as the sum of motions with three
independent frequencies, which are the relative pattern speed
$\omega_p$, and two others which correspond in the epicyclic limit to
$\omega$ and $\kappa$; on a loop, the action corresponding to the last
of these, the free radial oscillation, is zero.  In this case, we must
find the loops numerically.  Clearly, the loop concept can be extended
to a potential with 3 or more periodically varying components.
    
\section{LOOPS IN A REALISTIC GALAXY POTENTIAL}
As an example of a realistic double-barred galaxy potential, we
modified the `standard' single-bar model used by Athanassoula (1992)
and Piner \etal (1995), consisting of an axisymmetric disk and bulge,
and an $n=2$ Ferrers bar.  The larger bar in our model is prolate with
axial ratio $b/a = 0.4$, the quadrupole moment is $Q_m =
2.25\times10^{10}$ M$_{\odot}$ kpc$^2$ and its semimajor axis is $a_B$
= 7 kpc; the radius of the Lagrangian point $L_1$ is $r_L = 1.2 a_B$.
The small bar is also an $n=2$ Ferrers bar, with its mass, linear
size, and Lagrangian radius all set at 60\% of the corresponding
quantities for the big bar.  The central density and core radius of
the bulge are adjusted so that the central density of the whole system
is $1.0\times10^{10}$ M$_{\odot}$ kpc$^{-2}$ and the mass within 10
kpc is $5\times10^{10}$ M$_{\odot}$.  Our potential has a somewhat
weaker bar than the standard model of Athanassoula and Piner \etal,
with half the quadrupole moment, because we want to compare our exact
loops with those predicted by the epicycle approximation.

In the epicyclic approximation, our small bar has inner Lindblad
resonances at 0.6 and 1.3 kpc; its corotation radius is at 4.9 kpc,
beyond the outer of the inner Lindblad resonances of the larger bar at
2.9 kpc. We solve the linearized equations of motion in the $\cos (2
\theta)$ component of this potential, to find the loops in the middle
column of Figure 1. The four innermost loops behave like the inner
$x_1$ orbits in the potential of the smaller bar alone; the inner two
are almost circular, and the next two follow the small bar as it
rotates inside the larger bar.  Particles following orbits trapped
around these loops should be able to support the figure of the small
bar.  Further out, as the influence of the large bar becomes stronger,
the loops change smoothly from approximately following the $x_1$
orbits in the small bar, to being the counterparts of the $x_2$ orbits
in the big bar. Beyond the inner Lindblad resonance of the large bar
at radius 2.9 kpc, the loops are elongated along the $X$ axis and
behave like $x_1$ orbits in the large bar. The small bar has little
influence at these radii, except close to its corotation point at
$R=4.9$ kpc. The epicyclic approach fails close to the Lindblad
resonances and the corotation point of either bar, since the
perturbations $R_1$ and $\theta_1$ become large.

When we return to the nonlinear equations of motion in the exact
potential, we can again seek one-dimensional closed curves which
transform into themselves after a time $2\pi/\omega_p$, when the two
bars return to the same relative position. In general, the frequency
$\omega_p$ of the two bar patterns relative to each other is not
commensurable with the frequency $\omega$ with which any individual
particle circumnavigates the loop; so the position and velocity of a
particle on a generic loop at successive times $2 N \pi/\omega_p$,
where $N = 1, 2, \ldots$, will fill the one-dimensional curve which is
the loop. In the rotating frame in which the large bar remains
parallel to the $X$ axis, we searched for loops by following a
particle that starts on the $Y$ axis at an instant when the bars are
aligned.  We restricted ourselves to symmetric loops, for which the
starting velocity along the $Y$ direction is zero, and varied the
starting $X$ velocity so as to converge on the loop, beginning from an
initial estimate given by the epicyclic approximation.  The left
column of Figure 1 shows one of our loop solutions, with the epicyclic
approximation for comparison. The dots show successive positions of a
particle launched near the loop and followed for 400 orbits; it stays
close to the loop, which is seen to be stable.  Although (just as in a
single bar) particle orbits become chaotic near strong resonances,
loops exist over most of the range in radius, suggesting that a
self-consistent double bar could be built from particles on orbits
trapped around them.

The central column of Figure 1 shows that many of the loops in our
doubly-barred potential which lie within the small bar (semimajor axis
$a_S = 4.2$ kpc) remain aligned with it; particles following orbits
trapped around these loops can support the figure of this bar.  If the
small bar rotated somewhat more rapidly, its corotation radius would
coincide with large bar's inner Lindblad resonance; Tagger \etal
(1987) have suggested that a secondary bar is most likely to form with
resonances in this relation.  In that case, loops which are confined
within the small bar would not be influenced by any of the resonances
of the big bar, and almost all of them would remain aligned with the
small bar as it rotates relative to the larger bar; this would be the
most favorable case for building a strong secondary bar.  Louis \&
Gerhard (1988) have shown that two bars which make up a
self-consistent double-bar system must distort slightly as they rotate
through each other; but since this deformation has the same
periodicity as the relative rotation of the bars, it will not
introduce any extra frequencies into the potential, and the loops
should remain qualitatively similar to those we calculate here.

\section{GAS FLOW IN DOUBLE BARS}
Lindblad \& Lindblad (1994), and Wada (1994), added a dissipative term
to the equations of motion in a single bar, linearized the equations
using the epicyclic approximation, and studied the closed periodic
orbits in this system.  With the dissipative term included, the
deviations from circular orbits are finite at the Lindblad resonances,
and the major axis of the closed orbits rotates smoothly through 90
degrees; but the epicyclic approximation still breaks down at
corotation.  Inside corotation, a surprising correspondence was found:
the pattern of closed orbits closely reflects the gas flows modeled
hydrodynamically --- the damped epicyclic equations give us a cheap
way to preview the hydrodynamic simulation.  These two papers used
slightly different versions of the frictional term, and their results
are substantially independent of the exact form chosen.

Unlike stars, gas clouds orbiting in a single bar potential can only
fill orbits which do not cross themselves or intersect each other: gas
clouds on self-crossing or intersecting orbits would collide, removing
the gas from these orbits. In a double bar, if loops which are
solutions of linearized equations of motion including dissipation
should cross themselves or intersect other loops {\it at any given
  time}, gas will be quickly swept out of these loops.  In regions
where orbits or loops crowd together, shocks in gas are expected,
which would be visible in a real galaxy as dark dust lanes.

In order to preview gas flows in a doubly-barred potential, 
we included in  our equations of motion a dissipative term 
\begin{equation}
{\bf F}_{\tiny \rm friction} = - 2 \lambda \ 
[ \ \dot{\bf r} \ - \ {\bf \Omega}(R) \times {\bf r} \ ] \ / R
\end{equation}
proportional to the difference between the particle's velocity and 
the speed on a circular orbit at that radius;
the constant $\lambda$ prescribes the strength of the friction. The results 
for $\lambda$=16 km s$^{-1}$, which at $R=$4 kpc is equal to the value used 
by Lindblad \& Lindblad, are presented in the right column of Figure 1. 
This force corresponds only to about 8\% of the circular velocity on the flat 
part of the rotation curve in our potential, so 
it is small in relation to other speeds in the problem. 

The flow breaks into two parts: the outer pattern is generated by the
big bar and is almost steady in the rotating frame, while the inner
one rotates along with the smaller bar. The shocks induced separately
by each bar are at times connected (the two lower panels on Fig.1)
while at other relative phases of the two bars, they are disconnected
(two upper panels). Connected shocks may speed up the gas inflow into
the galaxy center, which might give rise to a long-timescale
periodicity in the feeding of an active nucleus.  In the dissipative
epicyclic approximation, particles remain on closed orbits or periodic
loops, so we cannot use this scheme to estimate the rate of gas
inflow.  Depending on the relative location of resonances induced by
both bars, the loop patterns can interfere with each other in
different ways: so the arrangement of dust lanes or the distribution
of gas can be used as a diagnostic of the bar pattern speeds in
observed double-bar systems.

We thank Ortwin Gerhard and Peter Englmaier for helpful discussions,
and the NSF and NASA for financial support through grants AST93-20403
and NAGW-2796.  WM acknowledges the hospitality of the Wroc{\l}aw
University Observatory, where part of this work has been done.

\vspace{1cm}

FIGURE CAPTIONS:

\noindent {\bf Figure 1:}
Loops (sets of particles) followed for 4 relative positions of the
bars, in the rotating frame of the large bar, which lies along the
X-axis. {\it Left:} Exact loops (solid lines) remain close to the
epicyclic approximation (dashed).  The two crossed lines mark the
orientation of the large and the small bar and extend to the outer of
their inner Lindblad resonances.  The dots represent successive
positions of a generic particle that started close to the loop, and
remains trapped about it.  {\it Center:} In the same potential, loops
calculated in the epicyclic approximation (solid lines) for the same
set of bar orientations.  Dashed circles, from the outermost inwards,
mark corotation of the small bar at 4.9 kpc, the outer inner Lindblad
resonance of the large bar at 2.9 kpc, and the two inner Lindblad
resonances of the small bar at 1.3 and 0.6 kpc. {\it Right:} In the
same potential, and for the same bar angles, loops for the
`dissipative epicyclic' equations, with the dissipative coefficient
$\lambda$ = 16 \kms.
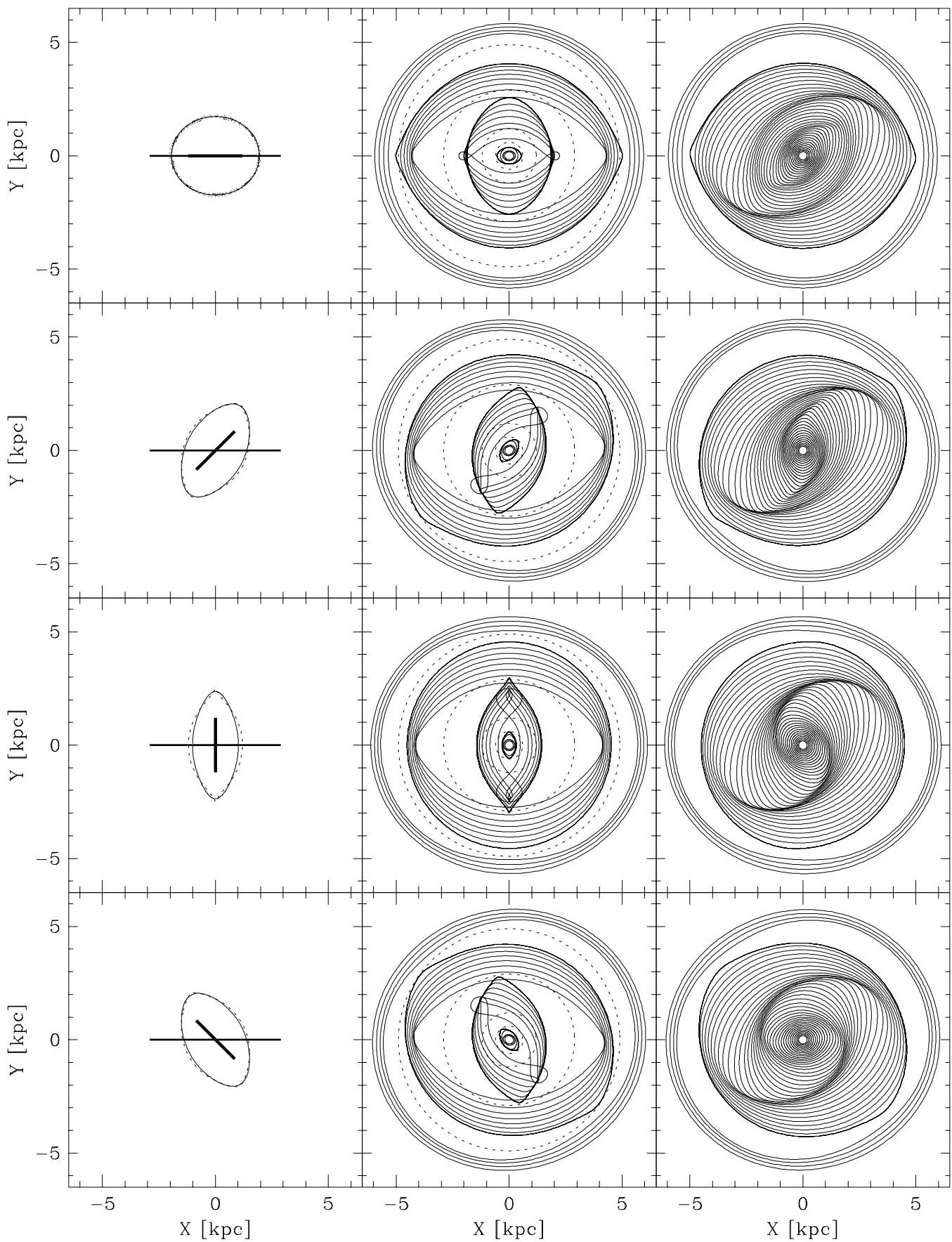
\end{document}